\documentclass[a4paper,11pt]{article}
\usepackage{pos}
\usepackage{bm}

\def\fire{{\textsc{Fire~6}}}
\def\Caravel{{\textsc{Caravel}}}
\newcommand{\mA}{m_{A}}
\newcommand{\mS}{m_{\phi}}
\newcommand{\ep}{\epsilon}
\newcommand{\bk}{\bm{k}}

\title{Numerical Unitarity for Binary Dynamics}
\author*{Fernando Febres Cordero}

\affiliation{Physics Department, Florida State University, Tallahassee, FL
32306-4350, USA}

\emailAdd{ffebres@hep.fsu.edu}

\abstract{
We present a calculation of the conservative two-body Hamiltonian of a compact
binary system including a spinning black hole.
We include up-to third order
corrections in Newton's constant $G$, all orders in velocity, and linear and
quadratic terms in spin. The results are obtained from the
classical limit of two-loop scattering amplitudes involving two massive scalars
and two massive spin-1 particles minimally coupled to gravity. We discuss the
usage of numerical techniques in our computation. In particular we show how the
numerical unitarity method is well suited to obtain results of relevance to the
physics program of current and future gravitational wave observatories.
}

\FullConference{16th International Symposium on Radiative Corrections:
Applications of Quantum Field Theory to Phenomenology ( RADCOR2023)\\
28th May - 2nd June, 2023\\
Crieff, Scotland, UK\\}

\begin{document}
\maketitle

\section{Introduction}
The first observation of gravitational waves by the LIGO-Virgo
collaboration~\cite{LIGOScientific:2016aoc} started a new era in the study of
the universe. Understanding the systems that produce those feeble spacetime
perturbations which we detect is critical to make the most of the experimental
programs of current and future (third-generation) gravitational wave
observatories. The gravitational wave signals for compact binary systems can be
described by three well-distinguished parts. First, the \textit{inspiral} where
the compact objects are very far apart, then the \textit{merger} where the
gravitational fields become strong (for example, when even horizons of
coalescent black holes touch), and finally the \textit{rigndown} where an
excited black hole evolves into a stable configuration.

Large amount of templates of these \textit{waveform} signals are required to
find gravitational wave events in the data sets collected by experiments. Though
nowadays numerical simulations of full evolution in general relativity are possible, they are rather computationally
expensive and so semi-analytic models are required to interpolate among simulations
in parameter space. One input that these models take are the conservative
potential of the compact binary systems during the inspiral phase. In this
presentation we focus on the calculation of those potentials as a
\textit{post-Minkowskian} (PM) expansion, that is as a series expansion in Newton's
coupling $G$. Great efforts have been devoted in recent years to compute
higher-order terms in the PM expansion. For example, for spinless systems,
third-PM corrections have been computed~\cite{Bern:2019nnu, Bern:2019crd, Kalin:2020fhe, DiVecchia:2021bdo, Herrmann:2021tct, Bjerrum-Bohr:2021din, Brandhuber:2021eyq} 
as well as fourth-PM corrections~\cite{Bern:2021dqo, Bern:2021yeh, Dlapa:2021npj, Dlapa:2021vgp}. 

The computation of PM corrections to systems involving spinning black holes is
of great interest as it is expected that they will play a key role in analyzing
a good fraction of detected gravitational waves.  In this presentation we focus
on the third-PM calculation of the conservative potential for a compact binary
system including a spinning black hole~\cite{FebresCordero:2022jts}. Other
recent activity include the calculation of the scattering angle, momentum
impulse and spin kick to fourth-PM order~\cite{Jakobsen:2023ndj, Jakobsen:2023hig} 
(see reference therein for further details). It is
expected~\cite{Buonanno:2022pgc} that with the increased sensitivity of
third-generation gravitational wave observatories up-to terms $\mathcal{O}(G^7)$
will be required to describe signals, and so the development of techniques that
can efficiently explore higher order correction is a necessity.

We report on our usage of numerical techniques
to compute the two-loop scattering amplitudes necessary for our calculation. In particular we discuss
the usage of the unitarity method~\cite{Bern:1994zx, Bern:1994cg, Bern:1995db, Bern:1997sc, Britto:2004nc}, 
in a numerical variant~\cite{Ita:2015tya, Abreu:2017idw, Abreu:2017xsl, Abreu:2017hqn, Abreu:2018jgq} 
which is well suited to deal with generic effective field theories, like for
example theories of gravity.

\section{Scattering Amplitudes from Numerical Unitarity}

We study the scattering process of a massive scalar particle with a massive
vector (spin-1) particle minimally coupled to gravity. This is described by the
Lagrangian:
\begin{align}
	\mathcal{L} ={}& \sqrt{-g}\left[-\frac{2 R}{\kappa^2}\ \ +\frac{1}{2}g^{\mu\nu}
	\partial_\mu\phi \partial_\nu\phi - \frac{1}{2}\mS^2\phi^2
	\ \ -\frac{1}{4}g^{\mu\alpha}g^{\nu\beta}F_{\alpha\beta}F_{\mu\nu} 
	+\frac{1}{2}m_A^2 g^{\mu\nu} A_\mu A_\nu\right]\,,
\label{eq:lagrangian}
\end{align}
where $\kappa=\sqrt{32\pi G}$ with $G$ Newton's constant, $\phi$ is a scalar
field, $A^\mu$ a vector field, $g_{\mu\nu}$ is the metric, $R$ the Ricci scalar,
and $F_{\mu\nu}=\partial_\mu A_\nu - \partial_\nu A_\mu$. We study the elastic
process $$A(p_1,\ep_1) + \phi(p_2) \to \phi(p_3) + A(p_4,\ep_4)\ ,$$ with $p_1^2
= p_4^2 = \mA^2$, $p_2^2 = p_3^2 = \mS^2$, and where $\ep_{1,4}$ are the
corresponding polarization vectors of the vector particles. We compute the
scattering amplitudes, for given polarization choices, numerically employing the
\Caravel{} framework~\cite{Abreu:2020xvt}. We do this over momentum
configurations with values on a particular number field, that of a finite field with a
large cardinality. This allows to employ these numeric evaluations to
reconstruct the associated analytic expressions (see
e.g.~\cite{vonManteuffel:2014ixa,Peraro:2016wsq}). The implementation of Feynman
rules from the Lagrangian above has been made with the help of the package
\textsc{xAct}~\cite{Brizuela:2008ra,Nutma:2013zea}.

\begin{figure}[ht]
\centering
\includegraphics[scale=0.6]{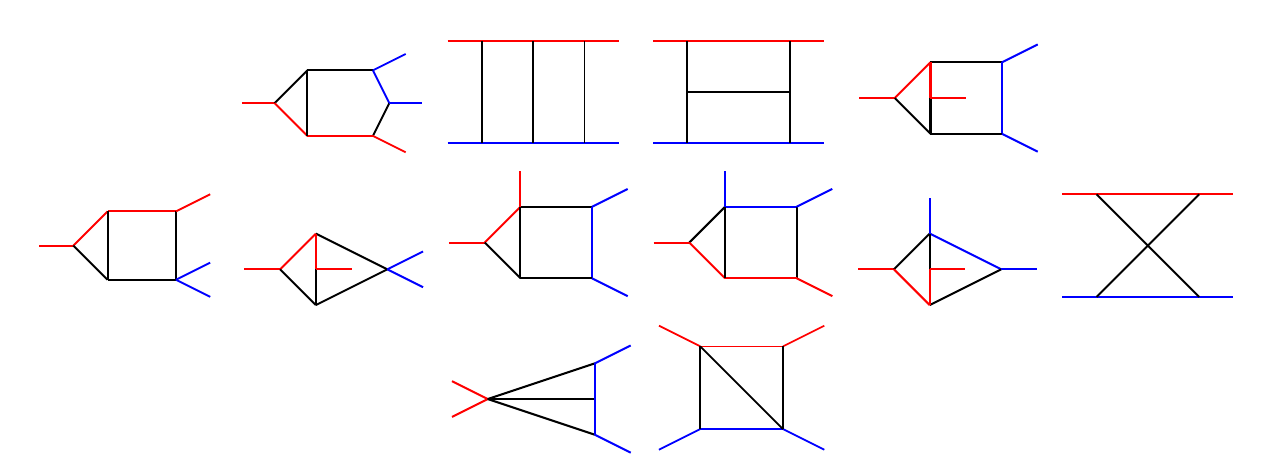}
\caption{Topologically inequivalent propagator structures contributing to the
classical two-body potential. Blue and red lines each represent a massive
particle, while black lines denote massless graviton exchanges.}
\label{fig:hierarchy}
\end{figure}

We employ the unitarity method~\cite{Bern:1994zx, Bern:1994cg, Bern:1995db, Bern:1997sc, Britto:2004nc} 
to compute the needed scattering amplitudes. We start by writing an ansatz for
the scattering amplitude:
$$
  \mathcal{M} =  \sum_{\Gamma \in \Delta} \sum_{i \in M_{\Gamma}}
         c_{\Gamma, i}\ \mathcal{I}_{\Gamma, i}\ ,
$$
where the sum is over all the master integrals $\{\mathcal{I}_{\Gamma, i}\}$,
here classified by a propagator structure $\Gamma$ and a
set of master integral indices $M_\Gamma$. In the
unitarity method we exploit analytic properties of the scattering amplitudes to
extract directly the coefficients $\{c_{\Gamma, i}\}$. 
Furthermore, in an approach well suited for numerical calculations, we introduce
an ansatz of the amplitude's integrand
$\mathcal{M}(\ell_l)$ 
as
\begin{equation}
 \mathcal{M}(\ell_l) = \sum_{\Gamma} \sum_{k\in Q_\Gamma} c_{\Gamma,k} 
 \frac{m_{\Gamma,k}(\ell_l)}{\prod_{j\in P_\Gamma} \rho_j}\ .
 \label{eqn:ansatz}
\end{equation}
The outer sum runs over all propagator structures $\Gamma$ encountered in the amplitude.
Given our interest in classical effects, the set of propagator structures is
considerably reduced with respect to the full quantum
amplitude~\cite{Bern:2019nnu,Bern:2019crd}. In Fig.~\ref{fig:hierarchy} we show all required
structures at the two-loop order. The $\rho_j$ are inverse propagators present
in $\Gamma$, and the functions $\{m_{\Gamma,k}(\ell_l)\}$ parametrize all
integrand insertions (up to a given power counting in loop momenta).
Finally, the coefficients $c_{\Gamma,k}$ contain all the process-specific
information and are functions of the external kinematics and the dimensional
regularization~\cite{tHooft:1972tcz} parameter $\epsilon=(4-D)/2$.

The family of functions $\{m_{\Gamma,k}(\ell_l)\}$ is labelled by a set of
indices $Q_\Gamma$. In principle, any complete set (up to a given power counting
in the loop momenta) of linearly independent functions can be employed for the
ansatz. Typically we use three types of sets. Consider the \textit{adaptive}
loop-momentum parametrization for a given propagator structure $\Gamma$:
\begin{equation}
\ell_l = \sum_{j\in B_l^p} v_l^j r^{lj} +
         \sum_{j\in B_l^t} u_l^j \alpha^{lj} +
         \sum_{i\in B^{ct}} \frac{n^i}{(n^i)^2} \alpha^{li} +
         \sum_{k\in B^\ep} n^k \mu_l^{k}\ ,
\end{equation}
where we have split the $D$-dimensional Minkowski space into four pieces. First,
$B_l^p$ represents the scattering plane spanned by external momenta connected to
$\ell_l$. Second, $B^{ct}$ is the so-called common-transverse space, the part of
the 4-dimensional Minkowski space transverse to all external momenta attached to
the propagator structure $\Gamma$. Then $B_l^t$ is the missing transverse piece
to complete the 4-dimensional Minkowski space with the two previous subspaces,
and finally we introduce a parametrization of the $\ep$-dimensional space with
$B^\ep$. The vectors $v_l^j$, $u_l^j$, $n^i$, and $n^k$, span their
corresponding spaces, and the variables left are the corresponding parameters to
characterize the loop momenta. In particular the $r^{lj}$ and $\mu_l^{k}$ can be
associated to inverse propagators of $\Gamma$.

Then when parametrizing a given integrand for a propagator structure $\Gamma$ we
can construct the bases:
\begin{enumerate}
  \item Tensor basis: where we construct  from all monomials (up-to
corresponding power counting) of the type
  $(\alpha^{lj})^{\vec a}(\alpha^{li})^{\vec b}$ with $j\in B_{l}^t$ and $i\in
B^{ct}$. The vectors $\vec a$ and $\vec b$ are the non-negative integer
exponents of the monomials.
  \item Scattering-plane tensor basis: starting from the tensor basis before, we
replace all monomials containing variables in $B^{ct}$ by corresponding
functions which integrate to zero using one-loop-like surface terms (see
e.g.~\cite{Abreu:2017xsl}).
  \item Master-surface basis: through the usage of unitarity-compatible
integration-by-parts (IBP) relations~\cite{Gluza:2010ws} one can further reduce
the left-over monomials of the previous basis by \textit{surface
terms}~\cite{Ita:2015tya} in such way that $Q_\Gamma=M_\Gamma\cup S_\Gamma$
with:
\begin{equation}
\int  \frac{d^D\ell_1d^D\ell_2}{(2\pi)^{2D}}\,
\frac{m_{\Gamma,i}(\ell_l)}{\prod_{k\in P_\Gamma} \rho_k} = \left\{
    \begin{array}{cc} I_{\Gamma,i} &  \mbox{for}\quad i\in
M_\Gamma\ \ \mathrm{\textcolor{black}{\small(master)}} \\ 0&  \mbox{for}\quad
i\in S_\Gamma\ \ \ \mathrm{\textcolor{black}{\small(surface)}} \end{array}\right.
\end{equation}
\end{enumerate}
The latter basis is particularly powerful as it trivializes the map between our
amplitude integrand ansatz and the integrated form in terms of master integrals.
In \Caravel{} we have an automated approach to build the first two types of
integrand parametrization and we have collected several master-surface
parametrization for amplitudes of interest (see e.g.~\cite{Abreu:2020lyk} for
more details).

A key factorization property of the integrand of an scattering amplitudes in
field theory occurs when we take internal (loop) propagators to on-shell limits.
That is, when the inverse propagators go to zero. In this limit equation
(\ref{eqn:ansatz}) gives:
\begin{equation}
  \sum_{\text{states}} \prod_{i \in T_{\Gamma}} \mathcal{M}_i^{\text{tree}} 
  (\ell_l^{\Gamma}) =
  \sum_{\substack{\Gamma' \geq \Gamma \\ k \in Q_{\Gamma'}}}
  \frac{{c_{\Gamma',k}}\ m_{\Gamma', k}(\ell_l^{\Gamma})}{
        \prod_{j \in (P_{\Gamma'} / P_{\Gamma})} \rho_j (\ell_l^{\Gamma})
       }\ ,
\label{eq:cuteq}
\end{equation}
where the sum on the RHS over $\Gamma' \geq \Gamma$ means for all propagator structures
containing all propagators or more of $\Gamma$. The momenta $\ell_l^{\Gamma}$ is
such that all inverse propagators in $\Gamma$ vanish. In the LHS of this
equation we have the product of all tree-level amplitudes characterized by the
vertices of the diagram $\Gamma$. This important relation is called the
\textit{cut equation}~\cite{Abreu:2017idw} and is the one used to compute the
coefficients $\{c_{\Gamma,k}\}$ of an scattering amplitude. The process samples
multiple values of $\ell_l^{\Gamma}$ and through linear algebra techniques,
returns all needed coefficients. This is the core of the so-called numerical
unitarity method~\cite{Ita:2015tya, Abreu:2017idw, Abreu:2017xsl, Abreu:2017hqn, Abreu:2018jgq}.
 
Whenever we have scattering tensors in our integrand bases, we have employed
IBP identities produced with the \fire{}~\cite{Smirnov:2019qkx} program. The
resulting expression is expanded as a Laurent series in small momentum transfer
$q^2$ and we also expand the corresponding master integrals accordingly
(using~\cite{Parra-Martinez:2020dzs}). After functional reconstruction,  we obtain final analytic expressions for
our scattering amplitudes. We refer the reader to the appendices
of~\cite{FebresCordero:2022jts} for the corresponding analytic expressions. 

\section{Effective Field Theory and Classical Potential}

We extract the classical potential through effective field theory (EFT)
techniques~\cite{Neill:2013wsa,Cheung:2018wkq,Bern:2020buy}. For that we employ
the non-relativistic EFT described by the Lagrangian:
\begin{align}
 L_{\textrm{EFT}} &= 
 \int_{\bk} \hat{\phi}^\dagger(-\bk)\left( \mathrm{i}\partial_t - \sqrt{\bk^2 + \mS^2}\right)\hat{\phi}(\bk)
 + \int_{\bk} \hat{A}^{\dagger,i}(-\bk)\left( \mathrm{i}\partial_t - \sqrt{\bk^2 + \mA^2}\right)\hat{A}^i(\bk) \\
 &- \int_{\bk,\bk^\prime} \tilde{V}_{ij}(\bk,\bk^\prime)\hat{A}^{\dagger,i}(\bk^\prime)\hat{A}^j(\bk)\hat{\phi}^\dagger(-\bk^\prime)\hat{\phi}(-\bk)\;,\label{eq:EFTLagrangian}\nonumber
\end{align}
where the integration $\int_{\bk}$ is $\int \frac{d^3\bk}{(2\pi)^3}$ and the
non-local function $\tilde V_{ij}$ is the potential which we will obtain
according to matching conditions. The potential $\tilde V_{ij}$ is decomposed in
terms of spin operators, which will produce the different linear (spin-orbit)
and quadratic terms in the Lagrangian that we will extract from our calculation.

\begin{figure}[ht!]
\begin{equation*}
 \includegraphics[height=2.2cm]{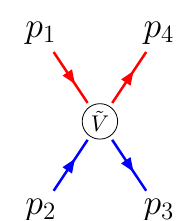} \raisebox{0.9cm}{\Large +}
 \includegraphics[height=2.2cm]{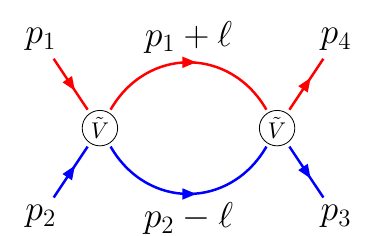}  \raisebox{0.9cm}{\Large +}
 \includegraphics[height=2.2cm]{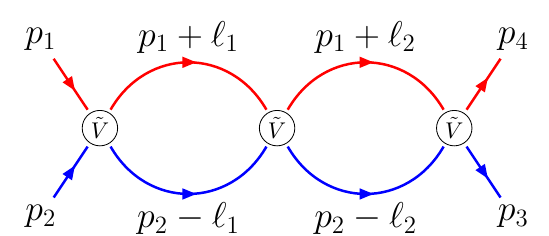}
\end{equation*}
 \caption{These iterated bubble diagrams give the corresponding amplitude in the
effective theory. The blue or red lines represent either the scalar or vector
particles.}
 \label{fig:eft_diags}
\end{figure}

The EFT amplitude is extracted from iterated bubble diagrams according
to~\cite{Cheung:2018wkq}. Furthermore the scattering amplitude, as well as the
potential $\tilde V_{ij}$ is expanded perturbatively in terms of $\kappa$. These
expansions are written explicitly in~\cite{FebresCordero:2022jts}. We
perform all calculations in the effective theory in dimensional regularization
around $3-2\epsilon$ dimensions. In this way all intermediate steps are properly
regularized and the matching procedure systematically removes infrared divergent
contributions to both the full theory amplitudes and to the effective theory
amplitudes. This is the first time this type of matching procedure has been
performed including all dimensional regularization contributions.

After matching the full theory amplitude to the EFT amplitude, order-by-order
and up to two loops, we obtained the conservative potential with contributions
up to third order in Newton's constant $G$. For brevity here we only include the
spin-orbit term of the potential. Its corresponding coefficient in momentum
space we write as $\tilde c_{L+1}^{(2)}$, where the $L$ represents the loop
order ($L$ is zero for the tree-level result, 1 for one loop, and 2 for two
loops). This is decomposed as:
	\newcommand{\Scalar}{\tilde{c}^{(1)}}
	\newcommand{\SpinOrbit}{\tilde{c}^{(2)}}
\begin{align}
	 \SpinOrbit_{L+1} &=\SpinOrbit_{L+1,\mathrm{red}} +\SpinOrbit_{L+1,\mathrm{iter}} +\frac{\Scalar_{L+1,\mathrm{red}}}{m_A^2(\gamma_1+1)}\ ,\label{eq:cecompositionCF}
\end{align}
where $\gamma_1 = E_A/\mA$. The coefficients $\Scalar_{L+1,\mathrm{red}}$ are
the ones appearing in the analogous spinless system. In the end, the full
expression for the spin-orbit coefficient up-to $\mathcal{O}(G^3)$ is:
\begin{align}
\Scalar_{1,\mathrm{red}}(\bk^2)={}&\frac{m_A^2m_\phi^2}{E^2\xi}\left(1-2\sigma^2\right)\,,\qquad
\Scalar_{2,\mathrm{red}}(\bk^2)={} \frac{3(m_\phi+m_A)m_\phi^2m_A^2}{4E^2\xi}(1-5\sigma^2)\,,\\
\Scalar_{3,\mathrm{red}}(\bk^2)={}&\frac{m_A^2m_\phi^2}{E^2\xi}\left[-\frac{2}{3} m_A m_\phi \left(\frac{\operatorname{arccosh}(\sigma)}{\sqrt{\sigma^2-1}} \left(-12 \sigma ^4+36 \sigma ^2+9\right)+22 \sigma ^3-19 \sigma \right)-2 (m_\phi^2+m_A^2) \left(6 \sigma ^2+1\right)\right]\nonumber\\
&+\frac{3Em_A^2m_\phi^2}{4E^2\xi}(m_A+m_\phi)\frac{(1-2\sigma^2)(1-5\sigma^2)}{(\sigma^2-1)}-\frac{3m_A^4m_\phi^4}{E^2\xi\bk^2}\ ,\\
\SpinOrbit_{1,\mathrm{red}}(\bk^2)={}&-\frac{2\sigma m_\phi}{E\xi}\ ,\qquad
\SpinOrbit_{2,\mathrm{red}}(\bk^2)={} \frac{\mS(4\mA+3\mS)\sigma(5\sigma^2-3)}{4E\xi(\sigma^2-1)}\ ,\\
\SpinOrbit_{3,\mathrm{red}}(\bk^2) ={}& \frac{m_\phi }{E\xi(\sigma^2{-}1)^2}\left[\left.{-}2m_A^2\sigma(3{-}12\sigma^2{+}10\sigma^4){-}\left(\tfrac{83}{6}{+ }27\sigma^2{-}52\sigma^4{+}\tfrac{44}{3}\sigma^6\right)m_Am_\phi- m_\phi^2\sigma\left(\tfrac{7}{2}{-}14\sigma^2{+}12\sigma^4\right)\right.\right.\nonumber\\
  &\qquad\qquad\qquad\left. {+}\frac{(4m_A{+}3m_\phi)E}{4}\sigma(2\sigma^2{-}1)(5\sigma^2{-}3){+}4m_A m_\phi\sigma(\sigma^2{-}6)(2\sigma^2{+}1)\sqrt{\sigma^2{-}1}\operatorname{arccosh}(\sigma)\right]\,,\\
\SpinOrbit_{1,\mathrm{iter}}(\bk^2)={}&0\,,\qquad
\SpinOrbit_{2,\mathrm{iter}}(\bk^2)={} E \xi  {\tilde c}_1^{\text{(2)}} \frac{\partial {\tilde c}_1^{\text{(1)}}}{\partial \bk^2}+{\tilde c}_1^{\text{(1)}} \left(E \xi  \frac{\partial {\tilde c}_1^{\text{(2)}}}{\partial \bk^2}+\frac{{\tilde c}_1^{\text{(2)}} \left(\frac{2 E^2 \xi }{\bk^2}+\frac{1}{\xi }-3\right)}{2 E}\right)\,,\\
\SpinOrbit_{3,\mathrm{iter}}(\bk^2)={}&\left({\tilde c}_1^{\text{(1)}}\right){}^2 \left({-}\frac{2}{3} E^2 \xi ^2 \frac{\partial ^2{\tilde c}_1^{\text{(2)}}}{\partial (\bk^2)^2}{+}\left(\xi  \left(3{-}\frac{E^2 \xi }{\bk^2}\right){-}1\right) \frac{\partial {\tilde c}_1^{\text{(2)}}}{\partial \bk^2}{+}{\tilde c}_1^{\text{(2)}} \left(\frac{\frac{1}{2 \xi }{-}2}{E^2}{+}\frac{3 \xi {-}1}{\bk^2}\right)\right)\\
	&{+}{\tilde c}_1^{\text{(1)}} \left({\tilde c}_1^{\text{(2)}} \left(\left({-}\frac{3 E^2 \xi ^2}{\bk^2}{+}6 \xi {-}2\right) \frac{\partial {\tilde c}_1^{\text{(1)}}}{\partial \bk^2}{-}\frac{4}{3} E^2 \xi ^2 \frac{\partial ^2{\tilde c}_1^{\text{(1)}}}{\partial (\bk^2)^2}\right)\right.\nonumber \\
	&\qquad\qquad\left. {+}\frac{4}{3} E \xi  \left(\frac{\partial {\tilde c}_2^{\text{(2)}}}{\partial \bk^2}{-}2 E \xi  \frac{\partial {\tilde c}_1^{\text{(1)}}}{\partial \bk^2} \frac{\partial {\tilde c}_1^{\text{(2)}}}{\partial \bk^2}\right){+}\frac{E^2 \xi ^2 \left({\tilde c}_1^{\text{(2)}}\right){}^2}{2 \bk^2}{+}{\tilde c}_2^{\text{(2)}} \left(\frac{\frac{2}{3 \xi }{-}2}{E}{+}\frac{E \xi }{\bk^2}\right)\right){-}\frac{1}{6} E^2 \xi ^2 \left({\tilde c}_1^{\text{(2)}}\right){}^3\nonumber\\
	&{+}{\tilde c}_1^{\text{(2)}} \left(\frac{2}{3} E \xi  \left(\frac{\partial {\tilde c}_2^{\text{(1)}}}{\partial \bk^2}{-}2 E \xi  \left(\frac{\partial {\tilde c}_1^{\text{(1)}}}{\partial \bk^2}\right){}^2\right){+}\frac{{\tilde c}_2^{\text{(1)}} \left(\frac{3 E^2 \xi }{\bk^2}{+}\frac{1}{\xi }{-}3\right)}{3 E}\right){+}\frac{2}{3} E \xi  {\tilde c}_2^{\text{(1)}} \frac{\partial {\tilde c}_1^{\text{(2)}}}{\partial \bk^2}{+}\frac{4}{3} E \xi  {\tilde c}_2^{\text{(2)}} \frac{\partial {\tilde c}_1^{\text{(1)}}}{\partial \bk^2}\nonumber\,,
\end{align}
where $\sigma = \frac{p_1\cdot p_2}{\mA\mS}$, $E = E_A + E_\phi$ and $\xi =
E_AE_\phi/E^2$. As mentioned above these coefficients are written in momentum
space. One can convert to position space if desired by a Fourier transform. We
have provided all corresponding expressions as ancillary files to
Ref.~\cite{FebresCordero:2022jts}.

We have performed a series of validation tests in our results. In particular our
calculation include spinless results which we have systematically compared to
the
literature~\cite{Holstein:2008sx,Bern:2019nnu,Bern:2019crd,Cristofoli:2020uzm,Kosmopoulos:2021zoq}.
Even more, we have compared to related results in the literature for spinning
observables or for post-Newtonian results and found
agreement~\cite{Levi:2015ixa,Levi:2015uxa,Levi:2017kzq,Vines:2017hyw,Vines:2018gqi,Liu:2021zxr,Jakobsen:2022fcj}.

\section{Conclusions}

We have presented a calculation including up-to third-order corrections in the
Newton's constant of the conservative potential for a compact binary system
including a spinning black hole. The calculation has been performed to all
orders in velocity and including up to quadratic terms in spin. We have employed
the numerical unitarity method in order to extract analytic expressions for the
scattering amplitudes involving massive scalar and massive vector particles
minimally coupled to gravity. This framework is flexible enough to carry further
calculations of interest to the future of gravitational wave astronomy. In
particular it is possible to explore higher spin terms, finite-size effects, and even higher
loop corrections.

\acknowledgments
We thank M.~Kraus, G.~Lin, M.~S.~Ruf, and M.~Zeng for collaboration in the work
presented here.
This work is supported in part by the U.S. Department of Energy under grant
DE-SC0010102.

\end{document}